\documentclass[twocolumn, aps, prb]{revtex4}
\usepackage{graphicx}
\usepackage{amssymb}
\usepackage{amsmath}

\def\lsim{\lower -0.3ex \hbox{$<$} \kern -0.75em \lower 0.7ex \hbox{$\sim$}}
\def\gsim{\lower -0.3ex \hbox{$>$} \kern -0.75em \lower 0.7ex \hbox{$\sim$}}

\newcommand{\GVec}[1]{\mbox{\boldmath$#1$}}

\def\mb{\bf}
\def\Vec#1{{\bf #1}}
\def\GVec#1{\mbox{\boldmath $#1$}}

\def\vare{\varepsilon}

\def\partd#1#2{\frac{\partial #1}{\partial #2}}

\begin{document}

\title{Orbital diamagnetism in multilayer graphenes:
Systematic study with the effective mass approximation}
\author{Mikito Koshino and Tsuneya Ando}
\affiliation{
Department of Physics, Tokyo Institute of Technology\\
2-12-1 Ookayama, Meguro-ku, Tokyo 152-8551, Japan}
\date{\today}

\begin{abstract}
We present a theoretical study 
on the orbital magnetism in multilayer graphenes
within the effective mass approximation.
The Hamiltonian and thus susceptibility can be decomposed into contributions from sub-systems
equivalent to monolayer or bilayer graphene.
The monolayer-type subband exists only in odd layers
and exhibits a delta-function susceptibility at $\vare_F=0$.
The bilayer-type subband appearing in every layer number gives a singular structure in the vicinity of $\vare_F=0$ due to the trigonal warping as well as a logarithmic tail away from $\vare_F=0$.
The integral of the susceptibility over energy is approximately given only by the layer number.
\end{abstract}

\maketitle

\section{Introduction}

Recently unconventional electronic properties 
of mono-crystalline graphenes attracts much attention
motivated by experimental fabrication,\cite{Novo04,Novo05,Zhan05-2}
although they were already the subject of theoretically study prior to the fabrication.\cite{McCl56,DiVincenzo_and_Mele_1984a,Gonzalez_et_al_1994a,Gonzalez_et_al_1996a,Shon_and_Ando_1998a,Gonzalez_et_al_2001a,Zheng_and_Ando_2002a,Ando_et_al_2002a,Suzuura_and_Ando_2002b,Ando_2005a}
Multilayer films which contain more than two layers
can also be synthesized,
and various phenomena depending on the layer number
have been reported.\cite{Novo05,Novo06,Ohta}
In this paper we present a theoretical study
on the orbital magnetism in multilayer graphenes.

The electronic structure of the monolayer graphene
is quite different from conventional metals,
because the conduction and valence bands touch 
at $K$ and $K'$ points in the Brillouin zone,
around which the dispersion becomes linear
like a relativistic particle.
In multi-layer graphenes,
the interlayer coupling makes a complex structure
around the band touching.
The electronic properties of graphene bilayer were
theoretically studied for the band structure \cite{McCa,McCa_gap}
and the transport properties.\cite{Koshino_and_Ando_2006a,Nils_multi,Nils_imps}
For few-layered graphenes of more than two stacks,
the electronic structure is investigated theoretically
in a $\Vec{k}\cdot\Vec{p}$ approximation,\cite{Guin}
a density functional calculation,\cite{Lati}
and a tight-binding model.\cite{Part,Liu}
On the experimental side, the band structures 
of graphenes from one to four layers
were recently measured using angle-resolved photoemission spectroscopy.\cite{Ohta}

The orbital magnetism in graphene-based systems was first studied
for a monolayer as a simple model to explain the large diamagnetism of
graphite.\cite{McCl56}
It was found that the susceptibility becomes
highly diamagnetic at $\vare=0$ (band touching point)
even though the density of states vanishes there.
The calculation was extended to graphite \cite{McCl57,Shar}
and to few-layered graphenes
as a model of graphite intercalation compounds.\cite{Safr,Safr84,Sait}
The Fermi surface of the graphite is known to be trigonally warped
around the band touching point \cite{McCl57}
and the effect of the warping on magnetization was discussed 
within the perturbational approach.\cite{Shar}
Recently, the disorder effects on 
the magnetic oscillation \cite{Gusy,Kosh}
and on the susceptibility \cite{Fuku07,Kosh}
were studied 
for the monolayer graphene.

Here we present a systematic study on the orbital magnetism 
for multilayer graphenes with arbitrary layer numbers
in the effective mass approximation.
We show that the Hamiltonian of a multilayer graphene
can be decomposed 
into those equivalent to monolayer or bilayer,
which allows us to study the dependence of
the susceptibility on layer numbers.
We take the trigonal warping effect into the calculation
and show that the fine structure around zero energy
gives rise to singular magnetic properties.

We introduce the model Hamiltonian and its decomposition into
subsystems in Sec.\ \ref{sec_form}, and present the calculation
of the magnetization in Sec.\ \ref{sec_mag}.
The discussion and summary are given in Sec.\ \ref{sec_disc}.

\section{Formulation}
\label{sec_form}

\begin{figure}
\begin{center}
\leavevmode\includegraphics[width=70mm]{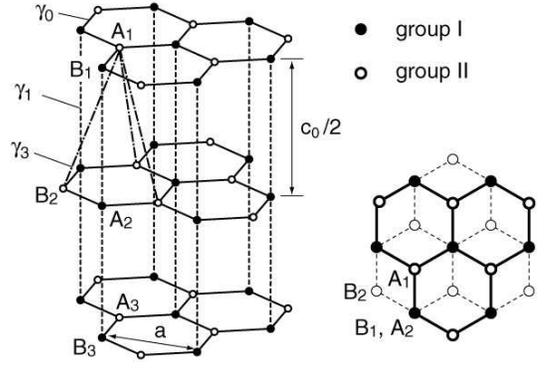}
\end{center}
\caption{
Atomic structure of multilayer graphene with
AB (Bernal) stacking.} 
\label{fig_schem}
\end{figure}

We consider a multilayer graphene
composed of $N$ layers of a carbon hexagonal network,
which are arranged in the AB (Bernal) stacking,
as shown in Fig.\ \ref{fig_schem}.
A unit cell contains $A_j$ and $B_j$ atoms on the layer $j = 1,\cdots,N$.
For convenience we divide carbon atoms into two groups as
\begin{eqnarray}
 {\rm Group\,\,I :} && B_1, \, A_2, \, B_3, \, \cdots \\
 {\rm Group\,\,II :} && A_1, \, B_2, \, A_3, \, \cdots
\end{eqnarray}
The atoms of group I are arranged along vertical columns
normal to the layer plane, 
while those in group II are above or below
the center of hexagons in the neighboring layers.
The lattice constant within a layer
is given by $a=0.246$ nm and the distance between adjacent layers
$c_0/2 = 0.334$ nm.

The system can be described by a {\bf k}$\cdot${\bf p} Hamiltonian
closely related to a three-dimensional (3D)
graphite model.\cite{Wall,Slon,McCl57,Dres} 
The low energy spectrum is given by the states 
in the vicinity of $K$ and $K'$
points in the Brillouin zone.
Let $|A_j\rangle$ and $|B_j\rangle$ be the Bloch functions at the $K$
point, corresponding to the $A$ and $B$ sublattices, respectively, of
layer $j$.
For monolayer graphene, the Hamiltonian around $K$ point 
for the basis  $|A_1\rangle$, $|B_1\rangle$ is written as
\cite{McCl56,DiVincenzo_and_Mele_1984a,Ando_2005a,Ajik,Kane_and_Mele_1997a}
\begin{eqnarray}
 H_0 = 
\begin{pmatrix}
 0 & \gamma k_- \\ \gamma k_+ & 0
\end{pmatrix},
\end{eqnarray}
where $k_\pm = k_x \pm i k_y$ and
\begin{equation}
\gamma = {\sqrt{3} \over 2} a \gamma_0,
\label{eq_gamma}
\end{equation}
with $\gamma_0$ being the nearest-neighbor coupling in a single layer.
We cite the experimental
estimation $\gamma_0 \approx 3.16$ eV.\cite{Toy} 

For the inter-layer coupling, we include parameters
$\gamma_1$, and $\gamma_3$, following the Hamiltonian
previously derived for a bilayer
graphene.\cite{McCa,Koshino_and_Ando_2006a} 
Here $\gamma_1$ represents 
the coupling between vertically neighboring atoms
in group I ($A_{2k} \leftrightarrow B_{2k\pm 1}$), and $\gamma_3$
between group II atoms on neighboring
layers ($B_{2k} \leftrightarrow A_{2k\pm 1}$), which are estimated to
$\gamma_1 \approx 0.39$ eV,\cite{Misu} 
and $\gamma_3 \approx 0.315$ eV.\cite{Doez}
If we look at the interaction between layers 1 and 2,
the matrix element $\langle A_2|{\cal H}|B_1\rangle$,
corresponding to the vertical bond, becomes $\gamma_1$
not accompanying in-plane Bloch number.
The matrix element $\langle B_2|{\cal H}|A_1\rangle$
is written as $\gamma' k_+$ with
\begin{equation}
\gamma' = {\sqrt{3} \over 2} a \gamma_3,
\end{equation}
similar to the intra-layer term $\langle A_1|{\cal H}|B_1\rangle$,
as the in-plane vector components from $A_1$ to $B_2$
are identical to those from $B_1$ to $A_1$.

Accordingly, if the basis is taken as $|A_1\rangle,|B_1\rangle$;
$|A_2\rangle,|B_2\rangle$; $\cdots$; $|A_N\rangle,|B_N\rangle$, the
Hamiltonian for the multilayer graphene
around the $K$ point becomes
\begin{eqnarray}
 {\cal H} =
\begin{pmatrix}
 H_0 & V & & & \\
 V^{\dagger} & H_0 & V^{\dagger}& & \\
  & V & H_0 & V & \\
  &  & \ddots & \ddots & \ddots
\end{pmatrix},
\label{eq_H}
\end{eqnarray}
with
\begin{eqnarray}
 V = 
\begin{pmatrix}
 0 & \gamma' k_+ \\ \gamma_1 & 0
\end{pmatrix}.
\label{eq_H0}
\end{eqnarray}
The effective Hamiltonian for $K'$ 
is obtained by exchanging
 $k_+$ and $k_-$ and replacing $\gamma_1$ with $-\gamma_1$.
The derivation of the effective Hamiltonian 
based on a tight-binding model is presented in Appendix \ref{sec_app_a}.

We show in the following that the Hamiltonian matrix (\ref{eq_H}) can be block-diagonalized into smaller matrices by choosing an appropriate basis independent of $\Vec{k}$.
First, we arrange the basis in the order of group I and then group II, i.e., $|B_1\rangle$, $|A_2\rangle$, $|B_3\rangle$, $\cdots$; $|A_1\rangle$, $|B_2\rangle$, $|A_3\rangle$, $\cdots$.
Then, Eq.\ (\ref{eq_H}) becomes
\begin{eqnarray}
 {\cal H} =
\begin{pmatrix}
{\cal H}_{11} & {\cal H}_{12} \\
{\cal H}_{12}^\dagger & {\cal H}_{22}
\end{pmatrix},
\end{eqnarray}
with ${\cal H}_{ij}$ being $N\times N$ matrices defined as
\begin{eqnarray}
 {\cal H}_{11} &=& 
\gamma_1\begin{pmatrix}
0 & 1 \\
1 & 0 & 1 \\ 
 & \ddots & \ddots & \ddots \\
 & & 1 & 0 & 1 \\
 & & & 1 & 0 
\end{pmatrix},\\
 {\cal H}_{12} &=& 
\gamma \begin{pmatrix}
k_+ &  \\
 & k_- \\
& & k_+ &  \\
& & & \ddots \\
& & & & k_\pm
\end{pmatrix},\\
{\cal H}_{22} &=&
\gamma' 
\begin{pmatrix}
0 & k_+ \\
k_- & 0 & k_- \\ 
& k_+ & 0 & k_+ \\ 
& & \ddots & \ddots & \ddots\\
& & & k_\mp & 0 & k_\mp \\
& & & & k_\pm & 0 
\end{pmatrix},
\end{eqnarray}
where the upper and lower signs correspond to odd and even $N$, respectively.

If we set $\Vec{k} = 0$ (the $K$ point), ${\cal H}_{12}$ and ${\cal H}_{22}$ vanish.
Remaining ${\cal H}_{11}$
is equivalent to the Hamiltonian 
of a one-dimensional tight-binding chain
with the nearest-neighbor coupling $\gamma_1$,
giving a set of eigenenergies
\begin{eqnarray}
\vare_{m} &=& \gamma_1 \lambda_{N,m} , \nonumber\\
\lambda_{N,m} &=& 2 \sin \frac{m\pi}{2(N+1)},
\label{eq_1D}
\end{eqnarray}
with 
\begin{eqnarray}
 m = -(N-1), \, -(N-3), \, \cdots , \, N-1. \label{eq_m}
\end{eqnarray}
Here, $m$ is an odd integer when the layer number $N$ is even, while $m$ is even when $N$ is odd, and therefore $m=0$ is allowed only for odd $N$.

The corresponding wave function is explicitly written as
\begin{equation}
 \psi_m(j) = \sqrt{\frac{2}{N+1}} \sin 
\left[\frac{(-m+N+1)\pi}{2(N+1)} j \right],
\label{eq_psi}
\end{equation}
where $\psi_m(j)$ represents the amplitudes at $|B_1\rangle$, $|A_2\rangle$, $|B_3\rangle$, $\cdots$ and satisfies
\begin{equation}
\sum_j \psi_m(j) \psi_{m'}(j) = \delta_{mm'} .
\label{eq_orthogonality}
\end{equation}
We have a relation between the wave functions $\psi_{m}$ and $\psi_{-m}$ as
\begin{equation}
 \psi_{-m}(j) = \psi_{m}(j)(-1)^{j+1}. 
\label{eq_bipartite}
\end{equation}

Now we construct the basis by assigning $\psi_{m}(j)$
to the atoms of group I and II as
\begin{eqnarray}
 |\phi_{m}^{\rm (I)}\rangle \!\! &=& \!\!
\psi_m(1) |B_1\rangle + \psi_m(2)|A_2\rangle + 
\psi_m(3)| B_3\rangle + \cdots \! , \quad \nonumber\\
\noalign{\vskip-0.150cm}
\\
\noalign{\vskip-0.250cm}
 |\phi_{m}^{\rm (II)}\rangle \!\! &=& \!\! 
\psi_m(1) |A_1\rangle + \psi_m(2)|B_2\rangle + 
\psi_m(3)| A_3\rangle + \cdots \! , \qquad \nonumber
\end{eqnarray}
and attempt to rewrite the Hamiltonian (\ref{eq_H}).
The matrix elements within group I come from ${\cal H}_{11}$
and become diagonal as is obvious from the definition,
\begin{eqnarray}
 \langle \phi_{m'}^{\rm (I)} |  {\cal H} | \phi_{m}^{\rm (I)} \rangle
&=& \delta_{m,m'} \gamma_1 \lambda_{N,m} .
\end{eqnarray}
Off-diagonal elements between 
$|\phi_{m}^{\rm (I)}\rangle$ and $|\phi_{m'}^{\rm (II)}\rangle$ are written 
from ${\cal H}_{12}$ as
\begin{eqnarray}
&& \!\!\!\!\! \langle \phi_{m'}^{\rm (II)} |  {\cal H} | \phi_{m}^{\rm (I)} \rangle \nonumber \\
&& = \gamma k_x  \sum_{j=1}^{N}  \psi_{m'}^*(j) \psi_{m}(j) + i \gamma k_y \sum_{j=1}^{N} \psi_{m'}^*(j) \psi_{m}(j) (-1)^j \nonumber \\
&& = \gamma(k_x \delta_{m,m'} - i k_y \delta _{m,-m'}).
\end{eqnarray}
In the second equality we used 
relation (\ref{eq_bipartite}) and
orthogonality (\ref{eq_orthogonality}).
Lastly, the matrix elements within group II are obtained
from ${\cal H}_{22}$ as
\begin{eqnarray}
& \!\!\!\! & \langle \phi_{m'}^{\rm (II)} |  {\cal H} | \phi_{m}^{\rm (II)} \rangle \nonumber \\
= & \!\!\!\! & \gamma' k_x  \sum_{j=1}^{N-1}  
\left[
\psi_{m'}^*(j+1) \psi_{m}(j)  + \psi_{m'}^*(j) \psi_{m}(j+1) 
\right]
\nonumber \\
+ \!\! & \!\!\!\! & \gamma' i k_y \sum_{j=1}^{N-1} (-1)^j 
\left[
\psi_{m'}^*(j+1) \psi_{m}(j) - \psi_{m'}^*(j) \psi_{m}(j+1) 
\right]
\nonumber\\
= & \!\!\!\! & \gamma' \lambda_{N,m}
(k_x \delta_{m,m'} + i k_y \delta _{m,-m'}).
\end{eqnarray}

The Hamiltonian is thus closed
in the subspace  
$\{|\phi_{m}^{\rm (I)}\rangle, \, |\phi_{-m}^{\rm (I)}\rangle, \, |\phi_{m}^{\rm (II)}\rangle, \, |\phi_{-m}^{\rm (II)}\rangle\}$
for each $|m|$.
Particularly, $m=0$ is special in that the subspace is spanned 
with only two bases
$\{|\phi_0^{\rm (II)}\rangle, \, |\phi_{0}^{\rm (I)}\rangle\}$,
while this is absent in even-layer graphenes.
The sub-matrix is written as
\begin{eqnarray}
 {\cal H}_{m=0} =
\begin{pmatrix}
0 & \gamma k_- \\
\gamma k_+ & 0 
\end{pmatrix},
\end{eqnarray}
which is independent of $\gamma_1$ and $\gamma_3$,
and equivalent to the Hamiltonian of the monolayer graphene.

For $m \neq 0$, we rearrange the basis as 
\begin{eqnarray}
\left\{
(|\phi_{m}^{\rm (II)}\rangle+|\phi_{-m}^{\rm (II)}\rangle)/\sqrt{2}, \quad
(|\phi_{m}^{\rm (I)}\rangle+|\phi_{-m}^{\rm (I)}\rangle)/\sqrt{2},
\right.
\nonumber \\
\left.
(|\phi_{m}^{\rm (I)}\rangle-|\phi_{-m}^{\rm (I)}\rangle)/\sqrt{2}, \quad 
(|\phi_{m}^{\rm (II)}\rangle-|\phi_{-m}^{\rm (II)}\rangle)/\sqrt{2} 
\right\},
\end{eqnarray}
where we take $m>0$ without loss of generality.
We then obtain
\begin{eqnarray}
 {\cal H}_{m} = 
\begin{pmatrix}
0 & \gamma k_- & 0 & \lambda \gamma' k_+ \\
\gamma k_+ & 0 & \lambda \gamma_1 & 0 \\
0 &  \lambda \gamma_1 & 0 & \gamma k_- \\
\lambda \gamma' k_- & 0 & \gamma k_+ & 0 
\end{pmatrix},
\label{eq_Hm}
\end{eqnarray}
with $\lambda = \lambda_{N,m}$.
This is equivalent to the Hamiltonian of a bilayer graphene except that
$\gamma_1$ and $\gamma'$ $(\propto\gamma_3)$ are multiplied by 
$\lambda$.

Thus the Hamiltonian of odd-layered graphene 
is composed of one monolayer-type and $(N-1)/2$ bilayer-type subbands
while that of even-layered graphene is composed of $N/2$ bilayers but
no monolayer.
The similar idea was previously
proposed for trilayer graphene without $\gamma_3$,
where it was shown that 
the energy spectrum becomes a superposition of that for a monolayer and 
for a bilayer.\cite{Guin}
Here we have extended this argument to 
decomposition of the Hamiltonian matrix,
and to systems with arbitrary number of layers including the trigonal warping.
We also note that $\Vec{k}$-independence of the basis
becomes important in the following sections, since this enables us to
write the magnetization as a sum over 
contributions from sub-Hamiltonians, which are independently calculated.

Many other parameters were introduced for the description of the band
structure of bulk graphite.\cite{McCl57,Dres,Shar,Char} 
The parameter $\gamma_4$
couples group I and II atoms sitting on the neighboring layers, such as
$A_j \leftrightarrow A_{j+1}$ or $B_j \leftrightarrow B_{j+1}$.
This
parameter does not change the qualitative feature of the low-energy
spectrum and therefore is not important.\cite{McCl57}
Parameters
$\gamma_2$ and $\gamma_5$ represent vertical hoppings between the
second-nearest neighboring layers for group II and I atoms,
respectively.  
Further, $\gamma_6$ is an energy difference between the
group I and II atoms due to difference in the chemical environment.
Inclusion of these parameters $\gamma_2$, $\gamma_5$, and $\gamma_6$
causes opening up of small energy gaps between the conduction and the
valence bands.  
However, these gaps do not play important roles in the
magnetization as will be discussed in the following.

In 3D limit, $N \rightarrow \infty$,
the eigenstate becomes a superposition
of opposite traveling waves with $\pm k_z$  along the stacking
direction.
The relation between the index $m$ and $|k_z|$ is obtained
by comparing the eigenenergy of ${\cal H}_{11}$, Eq.\ (\ref{eq_1D}),
to that of the 3D limit, $2\gamma_1 \cos(k_zc_0/2)$, as
\begin{equation}
{|k_z| c_0 \over 2} = {(-m+N+1)\pi \over 2(N+1)}.
\end{equation}

\begin{figure}
\begin{center}
\leavevmode\includegraphics[width=80mm]{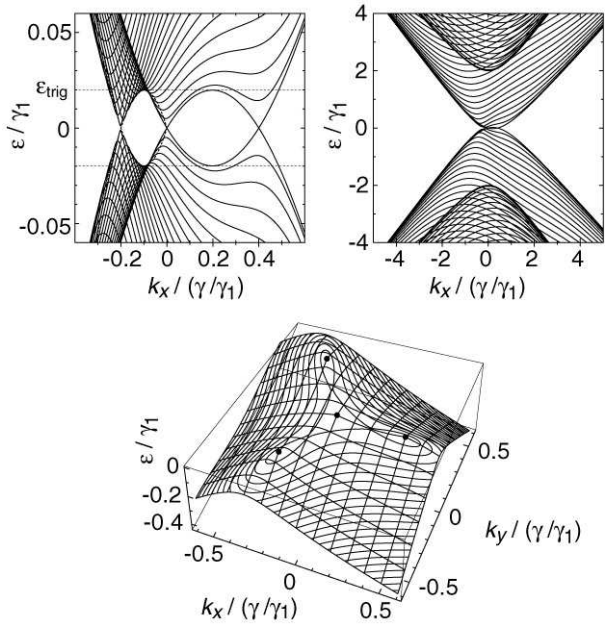}
\end{center}
\caption{
(Top) Projected band structure of the sub-Hamiltonian 
Eq. (\ref{eq_Hm})
with $\lambda = 2$ and $\gamma_3/\gamma_0=0.1$.
$|\vare| = \vare_{\rm trig} = 0.02\gamma_1$ 
is shown as horizontal dotted lines.
Right panel shows zoom out of the left.
(Bottom) 3D plot of the lower second band
around the band touching point.
Four Fermi points at $\vare=0$ indicated by dots.
}
\label{fig_band0}
\end{figure}

The band structure of the Hamiltonian 
(\ref{eq_Hm}) can be obtained by
replacing $\gamma_1$ by $\lambda\gamma_1$ and $\gamma_3$ by
$\lambda\gamma_3$
in that of the bilayer.\cite{McCa}
We plot in Fig.\ \ref{fig_band0} the dispersion 
for $\lambda = 2$, which has the maximum trigonal warping.
The middle two subbands stick together at $\vare =0$
while the remaining two bands appear only in the energy range 
$|\vare| > \lambda\gamma_1$.
If we neglect $\gamma_3$, the effective Hamiltonian 
for $|\varepsilon|\ll\lambda\gamma_1$ becomes
\begin{equation}
{\cal H} = {\hbar^2 \over 2m^*}
\begin{pmatrix}
0 & k_-^2 \\
k_+^2 & 0
\end{pmatrix}
,
\label{Eq:Bilayer_Low_Energy}
\end{equation}
which works for the reduced basis
\begin{equation}
\left\{
(|\phi_{m}^{\rm (II)}\rangle+|\phi_{-m}^{\rm (II)}\rangle)/\sqrt{2}, \quad
(|\phi_{m}^{\rm (II)}\rangle-|\phi_{-m}^{\rm (II)}\rangle)/\sqrt{2} 
\right\},
\end{equation}
giving a rotationally symmetric dispersion with the effective mass 
\begin{equation}
 m^* = \frac{\hbar^2 (\lambda\gamma_1)}{2\gamma^2}.
\label{eq_mstar}
\end{equation}

The term proportional to $\gamma_3$ is responsible for the trigonal warping effect,
which is most remarkable around the band sticking point $\varepsilon=0$.
Let us define
\begin{equation}
 \vare_{\rm trig} = \frac{1}{4}
\left(\frac{\lambda \gamma_3}{\gamma_0}\right)^2 
(\lambda\gamma_1).
\end{equation}
In the energy range $|\vare| \,\lsim\, \vare_{\rm trig}$, 
the Fermi line splits into four separated pockets,
one center part and three leg parts located trigonally,
which shrink into four Fermi points linearly with $\vare \rightarrow 0$.
We note that $\vare_{\rm trig}$ is proportional to $\lambda^3$
and thus very sensitive to $\lambda$,
while the energy of the higher-band bottom, $\lambda\gamma_1$, 
behaves linear to $\lambda$.
The maximum of $\lambda$ approaches 2 as the layer number
increases, so that  $\vare_{\rm trig}$ becomes as large as
$2(\gamma_3/\gamma_0)^2\gamma_1 \approx 8$ meV.

\begin{figure}
\begin{center}
\leavevmode\includegraphics[width=80mm]{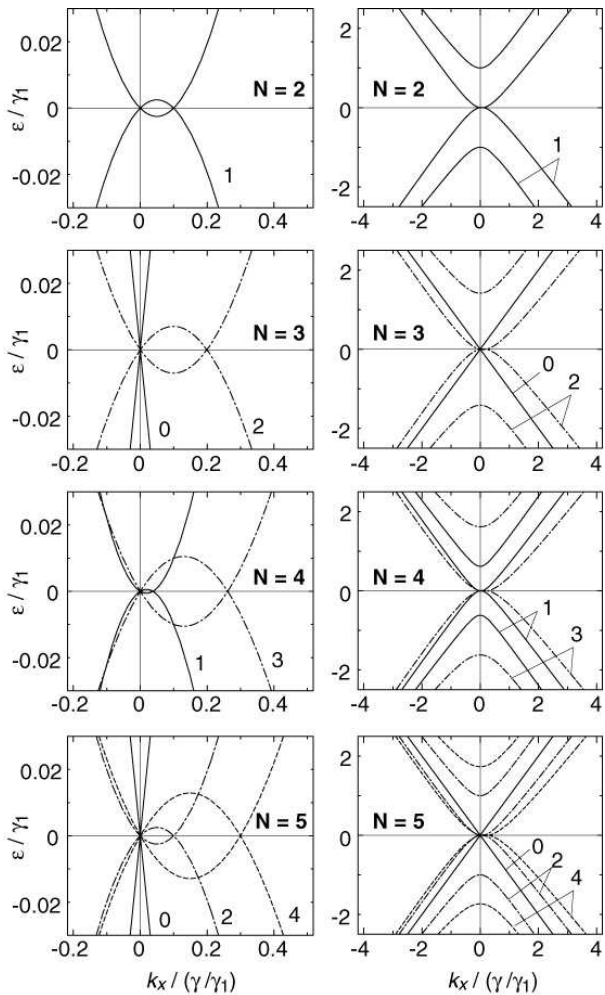}
\end{center}
\caption{
Band structures of multilayer graphenes,
$N=2,3,4,5$ with $\gamma_3/\gamma_0 = 0.1$, 
around the $K$ point (taken as origin)
along the $k_x$ axis.
Right panel shows a zoom-out of the left.
Numbers assigned to curves indicate $m$.
}
\label{fig_band}
\end{figure}

Figure \ref{fig_band} shows the band structures around the $K$ point
along the $k_x$ axis in the multilayer graphenes with $N=2$, 3, 4, and 5
and $\gamma_3/\gamma_0 = 0.1$.
The lists of $\lambda_{N,m}$ are given as
\begin{eqnarray}
N=1: && \{\lambda_{1,0} \} = \{ 0 \} \nonumber\\
N=2: && \{\lambda_{2,1} \} = \{ 1 \} \nonumber\\
N=3: && \{\lambda_{3,0},\,\lambda_{3,2} \} = \{0, \sqrt{2}\} \nonumber\\
N=4: && \{\lambda_{4,1},\,\lambda_{4,3} \} = 
\{(\sqrt{5}-1)/2, (\sqrt{5}+1)/2\} \nonumber\\
N=5: && \{\lambda_{5,0},\,\lambda_{5,2},\,\lambda_{5,4} \} 
= \{ 0, 1, \sqrt{3} \}.
\end{eqnarray}

While we have included $\gamma_0$, $\gamma_1$, and $\gamma_3$
in our graphene model, the extra parameter neglected here
may make some changes in the electronic structure.
The energy band of a few-layered graphene
has been calculated in the density functional calculation \cite{Lati}
and the tight-binding model.\cite{Part,Liu}
Those results differ from ours mainly in that
the band centers relatively shift depending on $m$, and that
a narrow gap opens where the conduction and valence
bands (within a single $m$) touch, 
and where different bands (with different $m$'s) cross.
Gaps are attributed to effects of couplings such as
$\gamma_2$, $\gamma_5$, and $\gamma_6$, which are
mentioned above.
In terms of the effective mass Hamiltonian (\ref{eq_H}),
those parameters appear as matrix elements without being multiplied by
the wave number $k_x$ and $k_y$, since they are associated with a
hopping along 
the $z$ axis
or a diagonal element.
Thus, they do not vanish at $k=0$ ($K$ or $K'$)
and lift the degeneracy to open a gap.
Apart from the gap opening,
the main feature of the trigonal warping is well described in the present model.
It should also be mentioned that 
an energy gap is induced
by an electric field perpendicular to the layer stacking direction,\cite{McCa_gap,Guin,Ohta,Cast}
where the electrostatic potential appears as matrix elements
independent of $k_x$ and $k_y$ as well.

\section{Magnetism of multilayer graphenes}
\label{sec_mag}

For the magnetic  susceptibility, we use the general expression
based on the linear response theory,\cite{Fuku71}
\begin{equation}
 \chi
= {\rm Im}  \int_{-\infty}^\infty d\vare f(\vare) F(\vare+i0),
\label{eq_chi}
\end{equation}
with
\begin{eqnarray}
F(z) = -\frac{g_{\rm v}g_{\rm s}}{2\pi L^2}\frac{e^2}{\hbar^2}
\sum_\Vec{k}{\rm tr}
\left(
G {\cal H}_xG {\cal H}_yG {\cal H}_xG {\cal H}_y
\right), 
\label{eq_F}
\end{eqnarray}
where $g_{\rm v}=2$ is the valley degeneracy, $g_{\rm s}=2$ is the spin degeneracy, 
and $L$ is the system size.
We defined here
${\cal H}_x = \partial {\cal H} /\partial k_x$,
${\cal H}_y = \partial {\cal H} /\partial k_y$,
$G(z) = (z-{\cal H})^{-1}$,
and $f(\vare)=[1+e^{(\vare-\mu)/k_B T}]^{-1}$
with the chemical potential $\mu$ and the temperature $T$.
The formula valid also for the Hamiltonian (\ref{Eq:Bilayer_Low_Energy}) is discussed in Appendix \ref{sec_app_b}.
By integration by parts in Eq.\ (\ref{eq_chi}), we have
\begin{eqnarray}
 \chi(T,\mu) &=& 
\int_{-\infty}^\infty d\vare 
\left(-\partd{f}{\vare}\right) \chi(0,\vare),
\label{eq_temp}
\end{eqnarray}
showing that the susceptibility at non-zero temperature is written
in terms of that at zero temperature.
The integration of $\chi$ over $\mu$ is independent of $T$.

We include the impurity scattering effects
by introducing a self-energy $-i\Gamma$ in the Green's function, i.e.,
$i0$ in (\ref{eq_chi}) is replaced by $i\Gamma$.
Here we simply assume the scattering rate $\Gamma = \hbar/2\tau$ 
to be independent of energy.

Using the decomposition of the Hamiltonian,
the magnetization of the $N$-layered graphene
can be written as a summation over
each sub-Hamiltonian.
The contribution from $m=0$ is exactly equivalent to
the susceptibility of a monolayer graphene,\cite{McCl56,Shar}
which becomes at zero temperature and in the clean limit,
\begin{equation}
  \chi_{\rm mono} =
- \frac{g_{\rm v}g_{\rm s}}{6\pi} \frac{e^2\gamma^2}{\hbar^2} \delta(\vare_F).
\label{eq_chi_mono}
\end{equation}
Thus, the odd-layer graphene always has a large diamagnetic peak at zero energy.
The delta-function dependence of $\chi_{\rm mono}$ agrees with the general property of the susceptibility in systems described by the $k$-linear Hamiltonian, as discussed in Sec.\ \ref{sec_disc}.

In the presence of disorder, the delta-function is broadened
into a Lorentzian with width $\Gamma$ and the same area,\cite{Fuku07} i.e.
\begin{equation}
  \chi_{\rm mono} =
- \frac{g_{\rm v}g_{\rm s}}{6\pi} \frac{e^2\gamma^2}{\hbar^2} {\Gamma\over \pi (\vare_F^2+\Gamma^2) },
\label{eq:Monolayer_with_Broadening}
\end{equation}
within the present model assuming a constant $\Gamma$.
The shape of the peak itself depends on the model disorder and we may have some different manner of broadening in a more realistic treatment.
In fact, in the monolayer graphene it was shown in a self-consistent Born approximation\cite{Shon_and_Ando_1998a} that $\chi$ has a much sharper peak at $\vare = 0$ than the Lorentzian and also a large tail proportional to $|\vare|^{-1}$ for $\vare\ne0$.\cite{Kosh}
In multi-layer cases effects of disorder are more complicated because of the presence of other bands.
This problem is out of the scope of this work.

The susceptibility of a bilayer graphene described by the Hamiltonian
(\ref{eq_Hm}) was analytically calculated for the case of
$\gamma_3=0$.\cite{Safr84}
The expression for $T=0$ and $\Gamma =0$ is given by
\begin{equation}
\chi = - \frac{g_{\rm v}g_{\rm s}}{4\pi}\frac{e^2\gamma^2}{\hbar^2}
\frac{\theta(\lambda \gamma_1 \!-\! |\vare_F|)}{\lambda \gamma_1}
\left(
\!-\! \ln\frac{|\vare_F|}{\lambda \gamma_1} - \frac{1}{3} 
\right) \!\! ,
\label{eq_chi_bi}
\end{equation}
with $\lambda = \lambda_{N,m}$,
where $\theta(t)$ is a step function defined by
\begin{equation}
\theta(t) = \left\{
\begin{array}{cc}
 1 & (t>0); \\
 0 & (t<0).
\end{array} \right.
\end{equation}
The susceptibility diverges logarithmically toward $\vare_F=0$, becomes
slightly positive for $|\vare_F| \,\lsim\, \lambda \gamma_1$, and
vanishes for $|\vare_F| > \lambda \gamma_1$ where the higher
subband enters.
In the presence of disorder, the logarithmic peak is broadened approximately as $\propto \ln\sqrt{\vare_F^2+\Gamma^2}$.

The integration of $\chi$ in Eq.\ (\ref{eq_chi_bi}) over the Fermi energy becomes $-(g_{\rm v}g_{\rm s}/3\pi)(e^2\gamma^2/\hbar^2)$ independent of $\gamma_1$, which is exactly twice as large as that of the monolayer graphene (\ref{eq_chi_mono}).
This arises due to the fact that the integral of $\chi$ over the Fermi energy is determined only by terms of the Hamiltonian matrix, proportional to $k_x$ or $k_y$, and is independent of terms independent of $k_x$ and $k_y$.
A proof of this important property is presented in Sec.\ \ref{sec_disc}.

\begin{figure}
\begin{center}
\leavevmode\includegraphics[width=80mm]{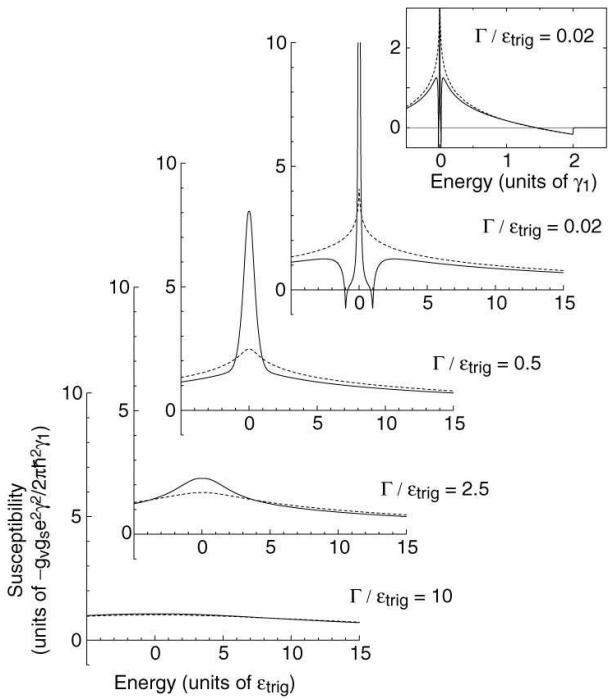}
\end{center}
\caption{
Susceptibility of the sub-Hamiltonian
(\ref{eq_Hm}) in the multilayer graphene
with $\lambda_{N,m}=2$, $\gamma_3/\gamma_0 = 0.1$
and several disorder strengths $\Gamma$.
Energy is scaled in units of $\vare_{\rm trig} = 0.02\gamma_1$.
Dashed curves show plots for $\gamma_3= 0$.
Inset at the top is a zoom out of the top panel 
($\Gamma = 0.02\vare_{\rm trig}$)
with units of energy $\gamma_1$.
}
\label{fig_chi}
\end{figure}

If we include the extra band parameter $\gamma_3$,
the low-energy structure of 
the susceptibility (\ref{eq_chi_bi}) drastically changes
due to the fine structure around the band touching point.
To demonstrate this, we numerically calculate $\chi$ for 
the Hamiltonian (\ref{eq_Hm})
in the case of the maximum trigonal warping, $\lambda_{N,m} = 2$.
Figure \ref{fig_chi} shows the susceptibility as a function
of $\vare_F$ with several values of $\Gamma$.
We take $\gamma_3/\gamma_1 = 0.1$, 
where the Fermi line splitting occurs in lower than
$\vare_{\rm trig} = 0.02\gamma_1$.
For reference we also plot the result without the trigonal warping, 
$\gamma_3 = 0$, as a dashed curve.

When we go down from high energy
in the top panel (the smallest $\Gamma$),
the susceptibility gradually deviates downward from
the logarithmic dependence of $\gamma_3 = 0$, 
and takes a sharp dip at $\vare = \vare_{\rm trig}$.
Remarkably we have a strong peak centered on $\vare=0$,
which is regarded as the effect of the linear dispersions 
around zero energy.
The integral of $\chi$ over the Fermi energy
is almost constant $ -(g_{\rm v}g_{\rm s}/3\pi)(e^2\gamma^2/\hbar^2)$
as discussed in Sec.\ \ref{sec_disc},
showing that the reduction in higher energies
compensates the zero energy peak.
As $\Gamma$ becomes larger, the peak begins to cancel 
with the reduction in high energy and 
the effect of $\gamma_3$ 
eventually disappears when $\Gamma \gg \vare_{\rm trig}$.
In contrast, the peak associated with the monolayer band $m=0$ becomes
broad in $\Gamma$ but never vanishes, as shown in Eq.\
(\ref{eq:Monolayer_with_Broadening}).

\begin{figure*}
\begin{center}
\leavevmode\includegraphics[width=170mm]{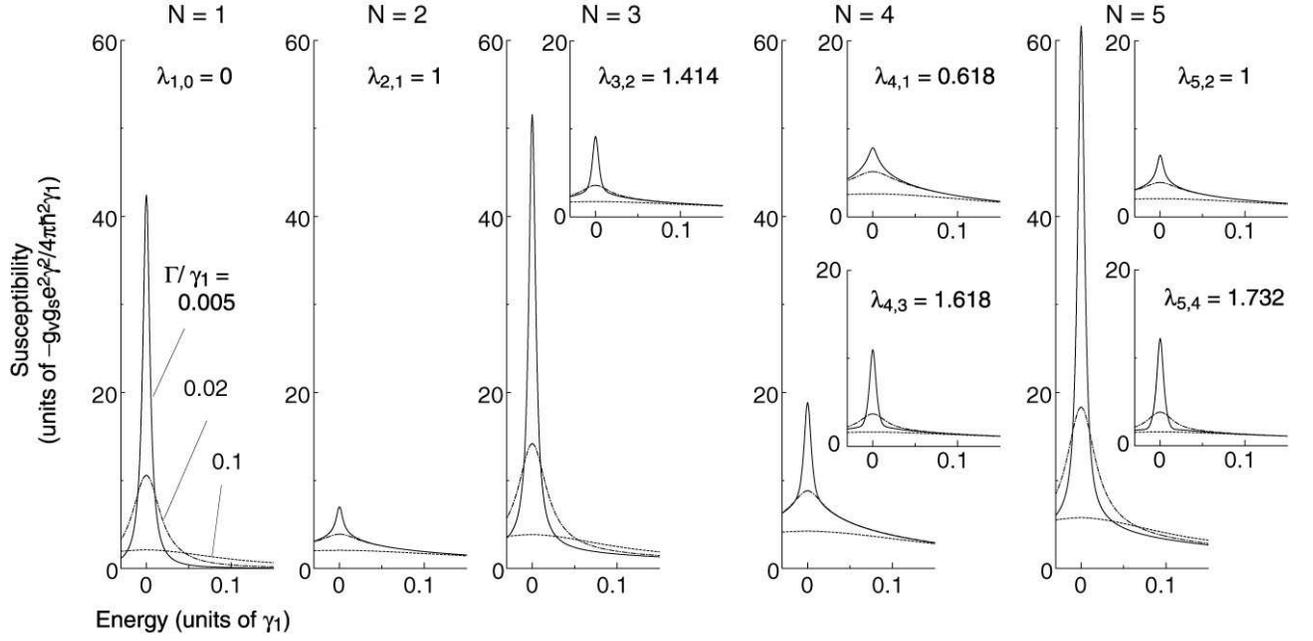}
\caption{
Susceptibility of multilayer graphenes with layer numbers $N=1$ to 5,
plotted against the Fermi energy.
Results shown for several disorder
 strengths specified by constant scattering rate $\Gamma$.
}
\label{fig_chi_multi}
\end{center}
\end{figure*}

Figure \ref{fig_chi_multi} shows $\chi(\vare_F)$ of graphenes with layer number from $N=1$ to 5 with several disorder strengths
$\Gamma$.
For $N\ge 3$, insets show the contributions 
from each of bilayer-type bands.
The result of odd $N$  always contains a monolayer-like component, 
which is exactly the same as $N=1$ and thus omitted in the inset.
We can see that odd-layered graphenes exhibit a particularly large peak,
which mainly comes from the monolayer-type band.
A bilayer-like component contains
a central peak due to the trigonal warping 
and a logarithmic tail in high energies,
in accordance with Fig.\ \ref{fig_chi}.

The layer-number dependence of the susceptibility in multilayer graphene has been studied for the graphite intercalation compounds.\cite{Sait}
This system can be viewed as independent multilayer graphenes bound by the intercalant layers, but the intercalants give a strong electrostatic potential along the stacking direction, leading to the charge redistribution among different layers.\cite{Ohno}
As a result the band structure and the magnetization are considerably
different from our system with a uniform electrostatic potential in the
vertical direction.

In isolated multilayer graphenes realized in recent experiments, we may
have some potential difference among layers depending on the
experimental environment, and this can also be tuned by the external
electric field as mentioned.
In Sec.\ \ref{sec_disc}, we will show that,
as long as the potential is not too strong
to alter the entire band structure,
this does not change the qualitative feature of the magnetization.

\section{Discussion}
\label{sec_disc}

The zero energy peak in the bilayer-type subband
originates in Dirac-like dispersions appearing around
four Fermi points.
Using the known results in a bilayer,\cite{McCa} we can show
that the sequence of the Landau levels in the center pocket
approximately becomes $\vare = {\rm sgn}(n) (\sqrt{2}\lambda_{N,m} \gamma'/l) \sqrt{|n|}$ with $N=0,1,2,\cdots$, and those in the three 
leg parts $\vare = {\rm sgn}(n) (\sqrt{6}\lambda_{N,m} \gamma'/l) \sqrt{|n|}$,
where $l=\sqrt{\hbar/(eB)}$ is the magnetic length.
Since the susceptibility is determined solely by Landau level energies,
we compare this to the monolayer's sequence
$\vare = {\rm sgn}(n) (\sqrt{2}\gamma/l) \sqrt{|n|}$
and obtain $\chi$ from each pocket by substituting 
$\gamma$ in Eq.\ (\ref{eq_chi_mono}).
We end up with
\begin{equation}
 \chi = 
10\left(\frac{\lambda_{N,m}\gamma_3}{\gamma_0}\right)^2
\chi_{\rm mono},
\label{eq_chi_fine}
\end{equation}
except for a constant coming from 
the integral over the lower energy states.
The zero-energy peak 
in Fig.\ \ref{fig_chi} fits well to the Lorentzian
with width $\Gamma$ and the area of the delta-function (\ref{eq_chi_fine}),
as long as $\Gamma \,\lsim\, \vare_{\rm trig}$.

The factor attached to $\chi_{\rm mono}$ becomes as large as 0.4 
when $\lambda_{N,m} = 2$ and $\gamma_3/\gamma_0 = 0.1$,
and therefore 
the singularity is not too small compared with that of the monolayer.
For $N$-layered graphene, a simple relation
\begin{equation}
\sum_{\lambda_{N,m} > 0} (\lambda_{N,m})^2 = N-1 
\end{equation}
leads to
the summation of (\ref{eq_chi_fine}) over all the bilayer-type subbands,
\begin{equation}
 \chi = 
10(N-1)\left(\frac{\gamma_3}{\gamma_0}\right)^2
\chi_{\rm mono} \approx 0.1\times(N-1)\chi_{\rm mono}.
\end{equation}
In Fig.\ \ref{fig_chi_multi} the peak height
becomes a little larger than this estimation
due to mixing with the logarithmic tail.


The delta-function dependence of $\chi$ in monolayer graphene
is a characteristic property common to general $k$-linear Hamiltonian.
This can be shown using the scaling argument.
We consider a Hamiltonian ${\cal H}$ which contains 
only terms linear in $k_x$ and $k_y$.
We change the energy and wave number scales
by an arbitrary factor $\alpha$ as
\begin{equation}
 \vare = \alpha \tilde\vare,
\quad
 k_i = \alpha \tilde k_i,
\end{equation}
then the Hamiltonian becomes formally identical 
under this transformation,
since the coefficients of $k$-linear terms in the Hamiltonian
remain unchanged.

Going back to the definition of $\chi$ in (\ref{eq_chi}) and (\ref{eq_F}),
$F(z)$ is scaled as
\begin{equation}
F(z) = \frac{1}{\alpha^2}\tilde F(\tilde z). 
\label{eq_F_scale}
\end{equation}
The function $F$ should depend only on the 
coefficients of $k$-linear terms and natural constants,
and thus is invariant under the scale transformation, namely 
we have $F = \tilde F$.
With (\ref{eq_F_scale}), we come up with a equation,
\begin{equation}
 F(z) = \frac{1}{\alpha^2}F\left(\frac{z}{\alpha}\right),
\end{equation}
which is satisfied solely by 
\begin{equation}
 F(z) = \frac{A}{z^2}.
\label{eq_F_explicit}
\end{equation}

A constant $A$ is related to the integral of the susceptibility 
$\chi(\vare_F)$ over the Fermi energy $\vare_F$.
From (\ref{eq_chi}), we generally have
\begin{equation}
\int_{-\infty}^{\infty} \! \chi(\vare)d\vare  
= -{\rm Im} \! \int_{-\infty}^{\infty} \! d\vare\,\vare F(\vare+i0)
= \frac{1}{2i}\oint_C \, dz \, z F(z),
\label{eq_chi_oint}
\end{equation}
where the integral path $C$ is a circle with an infinite radius
with anti-clockwise direction.
In the present system, 
(\ref{eq_F_explicit}) immediately gives the integral
as $\pi A$. 
This is an integral of the real function $\chi(\vare)$
and thus is real.
Substituting Eq. (\ref{eq_F_explicit}) with real $A$
in Eq. (\ref{eq_chi}), 
we finally obtain the explicit form of the zero-temperature susceptibility as
\begin{equation}
 \chi(\vare_F) = - {\rm Im} {A \over \varepsilon_F \!+\! i 0} = \pi A\delta(\vare_F).
\end{equation}


As discussed in Sec.\ \ref{sec_form}, 
the band structure in more realistic models
has an energy gap around zero energy
due to extra band parameters neglected in the present model.
It was also mentioned that the external 
electric field along stacking direction
opens an energy gap.
One might think that the gap would strongly reduce
the large diamagnetism at the band touching point.
However, we can show within the effective mass approximation
that the integral of susceptibility
over $\vare_F$ is independent of any kind of matrix elements
without $k_x$ and $k_y$, which are responsible for gap opening.
This is obvious from the general expression (\ref{eq_chi_oint});
even if the Hamiltonian contains $k$-independent terms
in addition to $k$-linear terms, 
they can be safely neglected in the integral
as they are infinitesimal compared to $|z|$ on the path $C$.
In the effective-mass model of the multilayer graphene we immediately conclude that the integral is independent of $\gamma_1$, $\gamma_2$, $\gamma_5$, $\gamma_6$, and any other parameters independent of the wave vector.
Thus we expect that the large diamagnetic peak is still visible
even when a gap opens, while it may get broadened in energy by the gap
width.
The diamagnetism in narrow gap systems is known in bismuth\cite{Fuku70} and recently studied for the gapped Dirac fermion.\cite{Naka}
Any further discussion requires a direct computation of the magnetization including extra parameters, but we leave this for the future study.

The integral of the susceptibility can be calculated
by the Hamiltonian with the $k$-independent terms dropped,
and thus depends only on the band parameters associated with
$k$-linear terms.
In our model (\ref{eq_H}),
the value is mainly determined by the dominant parameter $\gamma_0$,
while $\gamma_3$ gives a correction at most
of the order of $(\gamma_3/\gamma_0)^2 \sim 0.01$.
The correction must be the second order in $\gamma_3$
because we can change $\gamma_3$ to $-\gamma_3$ in the Hamiltonian
with a unitary transformation
multiplying the base on layer $j$ by $(-1)^j$.
As a result, 
the integral of $\chi(\vare_F)$ for the bilayer-type Hamiltonian 
becomes almost twice as large as the monolayer's,
and the summation over all the subsystems in $N$-layered graphene
becomes approximately $N$ times as large as the monolayer's.

It is instructive to derive the susceptibility starting from the Landau-level energies.
In the monolayer graphene, the thermodynamic function $\Omega$ is given by
%
\begin{eqnarray}
&& \Omega = - k_{\rm B} T g_{\rm v} g_{\rm s} {1\over 2\pi l^2} \sum_n g(\varepsilon_n) \varphi(\varepsilon_n) , \\
&& \varphi(\varepsilon) = \ln \big\{ 1 \!+\! \exp[\beta(\mu \!-\! \varepsilon)] \big\} ,
\end{eqnarray}
%
where $\beta=1/k_BT$, $\varepsilon_n={\rm sgn}(n)\hbar\omega_B\sqrt{|n|}$ with $\hbar\omega_B=\sqrt2\gamma/l$, and $g(\vare)$ a cutoff function which gradually decays to zero for $|\vare| \,\gsim\, \varepsilon_c$ with cutoff energy $\varepsilon_c$.
We can rewrite this as
%
\begin{equation}
\Omega = - k_{\rm B} T g_{\rm v} g_{\rm s} {1\over 2\pi l^2} \sum_{n=0}^\infty \Big( 1 \!-\! {1\over 2} \delta_{n0} \Big) H(nh) ,
\end{equation}
%
where
%
\begin{equation}
H(x) = g(\sqrt{x}) \ln \big[ 1 \!+\! 2 \exp(\beta\mu) \cosh(\beta\sqrt{x}) \!+\! \exp(2\beta\mu) \big] ,
\end{equation}
%
with $h=(\hbar\omega_B)^2$.
\par
%
Expanding the integral
%
\begin{equation}
\int_0^\infty \! H(x) d x = \int_0^{h/2} \! H(x) dx + \sum_{j=1}^\infty \int_{-h/2}^{h/2} \! H( x \!+\! h j) dx ,
\end{equation}
%
with respect to $h$, we immediately have
%
\begin{eqnarray}
& \!\!\!\!\!\! & h \Big[ {1\over 2} H(0) + \sum_{j=1}^\infty H( x \!+\! h j) \Big] \nonumber \\
= & \!\!\!\!\!\! & \int_0^\infty H(x) d x - {1\over 12} h^2 \Big[ H'(0) \!+\! {1\over 2} H'(\infty) \Big] , \
\end{eqnarray}
%
up to the second order in $h$ or in $B$.
Then, we have
%
\begin{equation}
\Omega = \Omega_0 + \Delta\Omega ,
\end{equation}
%
where $\Omega_0$ is the thermodynamic function in the absence of a magnetic field and
%
\begin{eqnarray}
\Delta \Omega &=& {1\over 12} { g_{\rm v} g_{\rm s} (\hbar\omega_B)^2 \over 2\pi l^2 } { \beta \exp(\beta\zeta) \over [ 1 \!+\! \exp(\beta\zeta) ]^2 } \nonumber\\
&=& {g_{\rm v} g_{\rm s} \gamma^2 \over 12 \pi l^4} \int_{-\infty}^{\infty} \Big( \!-\! {\partial f(\varepsilon) \over \partial \varepsilon } \Big) \delta(\varepsilon) d \varepsilon .
\label{Eq:Thermodynamic_potential:_Monolayer_Graphene}
\end{eqnarray}
%
Applying the relation $\Delta\Omega=\chi B^2/2$, we obtain $\chi_{\rm mono}$ given by (\ref{eq_chi_mono}) at zero temperature.
\par
%
We should note that the thermodynamic function in the absence of a magnetic field is given by
%
\begin{equation}
\Omega_0 = - k_{\rm B} T g_{\rm v} g_{\rm s} {1\over 2\pi l^2} \sum_n \int_{-1/2}^{1/2} g(\varepsilon_{n+t}) \varphi(\varepsilon_{n+t})  dt .
\end{equation}
%
For contributions of states with $|n|\gg1$, we can expand the above with respect to $t$ and have to the lowest order in the field strength $B$
%
\begin{eqnarray}
\Delta \Omega & \! = \! & {g_{\rm v} g_{\rm s} \over 2\pi l^2 } {1\over 96} (\hbar\omega_B)^4 \sum_{|n|\gg0} [ \varepsilon_n^{-3} f(\varepsilon_n) \!-\! \varepsilon_n^{-2} f'(\varepsilon_n) ] \nonumber\\
& \! \approx \! & {g_{\rm v} g_{\rm s} \over 2\pi l^2 } {(\hbar\omega_B)^2 \over 48} \!\! \lim_{\delta\rightarrow+0} \Big( \int_\delta^\infty [ \varepsilon^{-2} f(\varepsilon) \!-\! \varepsilon^{-1} f'(\varepsilon) ] d \varepsilon \nonumber \\
&& \qquad\qquad\qquad\quad  - \int_{-\infty}^{-\delta} [ \varepsilon^{-2} f(\varepsilon) \!-\! \varepsilon^{-1} f'(\varepsilon) ] d \varepsilon \Big) \nonumber \\
& \! = \! & - {g_{\rm v} g_{\rm s} \over 2\pi l^2 } {1\over 24} (\hbar\omega_B)^2 \int_{-\infty}^\infty \Big( \!-\! {\partial f(\varepsilon) \over \partial \varepsilon} \Big) \delta(\varepsilon) d \varepsilon .
\end{eqnarray}
%
This gives a ``paramagnetic'' susceptibility.
For $n\!=\!0$, on the other hand, the change in the thermodynamic potential is calculated as
%
\begin{equation}
\Delta \Omega = {g_{\rm v} g_{\rm s} \over 2\pi l^2 } {1\over 8} (\hbar\omega_B)^2 \int_{-\infty}^\infty \Big( \!-\! {\partial f(\varepsilon) \over \partial \varepsilon} \Big) \delta(\varepsilon) d \varepsilon .
\end{equation}
%
The sum of these two contribution is the same as Eq.\ (\ref{Eq:Thermodynamic_potential:_Monolayer_Graphene}), as is expected.
\par
%
In the bilayer graphene, the Landau level with $\gamma_3=0$ in the region $|\vare| \ll \gamma_1$ can be calculated from the Hamiltonian (\ref{Eq:Bilayer_Low_Energy}) as \cite{McCa}
\begin{equation}
\vare_{sn} = s \hbar \omega_c \sqrt{n(n+1)} ,
\label{eq_LL}
\end{equation}
where $\omega_c = eB/m^*$ with $m^*$ defined in Eq.\ (\ref{eq_mstar}),
$s=\pm 1$, and $n=0,1,2,\cdots$.
We have doubly degenerate levels at zero energy ($n=0$, $s=\pm 1$),
while the spacing gradually becomes constant as $n$ goes higher.
In a similar but more complicated manner, the susceptibility is calculated as
\begin{equation}
\chi =
-\left(\frac{e\hbar}{2m^*}\right)^2
\frac{g_{\rm v}g_{\rm s} m^*}{2\pi\hbar^2}
\int_{-\infty}^\infty g(\vare) 
\ln \frac{\vare_c}{e |\vare|}
\left(-\frac{\partial f}{\partial \vare}\right)
d\vare,
\label{eq_chi_bi_alt}
\end{equation}
which correctly describes the logarithmic divergence
around zero energy in the rigorous expression 
(\ref{eq_chi_bi}), as expected.
A constant term independent of energy is missing in
Eq. (\ref{eq_chi_bi_alt})
since this depends on all the low-energy bands which are
neglected in this calculation.

We can understand the logarithmic dependence 
intuitively by looking into the Landau-level sequence.
The Landau level energy can be expanded
for large $n$ as
\begin{equation}
\vare_{sn} = s \hbar \omega_c \Big[
\Big(n + \frac{1}{2}\Big)
-\frac 1 8 \Big(n+\frac 1 2 \Big)^{-1} + \cdots
\Big],
\end{equation}
where the first term gives the constant interval,
and the second gives a shift toward zero energy, which is rewritten as
$-(\hbar\omega_c)^2/(8 \vare_{sn})$.
For $\varepsilon_F<0$, for example, the change in the total energy due to the energy shift is calculated as
\begin{equation}
\Delta E = - {g_{\rm v} g_{\rm s} m^* \over 2\pi\hbar^2} \int_{-\varepsilon_c}^{\varepsilon_F} \!\! {(\hbar\omega_c)^2 \over 8\varepsilon} d\varepsilon 
= {B^2 \over 2} {g_{\rm v} g_{\rm s} \over 4\pi} {e^2\gamma^2 \over \hbar^2 \lambda \gamma_1} \ln {\varepsilon_c \over |\varepsilon_F|} ,
\end{equation}
giving the $\ln|\varepsilon_F|$ dependence of the susceptibility.

For the Hamiltonian (\ref{Eq:Bilayer_Low_Energy}) containing terms
proportional to $k_{\pm}^2$, the susceptibility 
formula (\ref{eq_chi}) with (\ref{eq_F}) is no longer
valid, since this was originally derived for systems
in which $x$ commutes with ${\cal H}_y$ and
 ${\cal H}_{yy}=\partial^2{\cal H}/\partial k_y^2$.
The modified formula should be
\begin{eqnarray}
F(z) & \!\! = \!\! & - {g_{\rm v} g_{\rm s} \over 4\pi L^2} {e^2\over \hbar^2} \sum_{\bf k} {\rm tr} \Big( G {\cal H}_x G {\cal H}_y G {\cal H}_x G {\cal H}_y \nonumber \\
\noalign{\vspace{-0.150cm}}
&& \qquad\qquad\qquad\enspace - 2 G {\cal H}_x G {\cal H}_x G {\cal H}_y G {\cal H}_y \nonumber \\
&& - {1\over 2} G {\cal H}_y G {\cal H}_{xx} G {\cal H}_y - {1\over 2} G
 {\cal H}_x G {\cal H}_{yy} G {\cal H}_x \Big). \quad
\label{eq_F_gen}
\end{eqnarray}
This is derived in Appendix \ref{sec_app_b}.
A scaling argument similar to the case of the
monolayer graphene then gives $F(z)\propto 1/z$.
This again leads to the logarithmic dependence 
of $\chi$ on the Fermi energy, which coincides with
(\ref{eq_chi_bi_alt}) apart from a constant.

\par
%


The experimental measurements of the magnetization of two-dimensional electron systems were performed on the semiconductor heterostructures, by using the superconducting quantum interference device (SQUID)\cite{Stor,Mein} or using the torque magnetometer.\cite{Eise,Pott,Wieg}
We expect that the detection of the graphene magnetism is also feasible with those techniques.

We have studied the orbital magnetism of multilayer graphene
with the Bernal stacking in the effective mass approximation.
We have demonstrated that
the Hamiltonian and thus the susceptibility can be decomposed into
those equivalent to the monolayer or bilayer bands.
The monolayer-like band 
exists only in odd-layered graphenes and gives a strong 
diamagnetic peak at $\vare_F =0$.
The bilayer-like bands 
always exist and present a strong diamagnetism
in the vicinity of zero energy,
unless the fine band structure caused by $\gamma_3$
is destroyed by the disorder.

\section*{ACKNOWLEDGMENTS}

This work has been supported in part by the 21st Century COE Program at
Tokyo Tech \lq\lq Nanometer-Scale Quantum Physics'' and by Grants-in-Aid
for Scientific Research from the Ministry of Education, 
Culture, Sports, Science and Technology, Japan.

\appendix

\section{Effective mass Hamiltonian}
\label{sec_app_a}
We derive in the following the effective mass equation 
Eq.\ (\ref{eq_H}) describing 
states in the vicinity of $K$ point in a multilayer graphene,
by starting from the one-orbital tight-binding model.
The following is nothing but a straightforward extension of the monolayer case.\cite{Ajik,Ando_2005a}
In a tight-binding model, the wave function is written as
\begin{eqnarray}
 \psi(\Vec{r}) = \sum_{j} 
\bigg[
\sum_{\Vec{R}_{A_j}} 
\psi_{A_j}(\GVec{\rho}_{A_j})
\phi(\Vec{r}-\Vec{R}_{A_j}) \quad
\nonumber\\
+ \sum_{\Vec{R}_{B_j}} 
\psi_{B_j}(\GVec{\rho}_{B_j})
\phi(\Vec{r}-\Vec{R}_{B_j}) 
\bigg],
\end{eqnarray}
where $j=1,2,\cdots,N$ is the layer index,
$\phi(\Vec{r})$ is the wave function of the $p_z$ orbital
of a carbon atom located at the origin,
as a function of three-dimensional position $\Vec{r}$.
$\Vec{R}_{X}$ is the three-dimensional position
of the site $X$, and $\GVec{\rho}_{X}$ is a two-dimensional
component of $\Vec{R}_X$, parallel to the layer.


In the model including hopping parameters
$\gamma_0$, $\gamma_1$, and $\gamma_3$ defined in Sec.\ \ref{sec_form},
Schr\"{o}dinger's equation can be written as follows:
For odd $j$,
\begin{eqnarray}
 \vare \psi_{A_{j}}(\GVec{\rho}_{A_{j}})
&=&
-\gamma_0 \sum_{l=1}^3  
\psi_{B_{j}}(\GVec{\rho}_{A_{j}} - \GVec{\tau}_l)
\nonumber\\
&&
\hspace{-20mm}
+\gamma_3 \sum_{l=1}^3  
\left[
\psi_{B_{j+1}}(\GVec{\rho}_{A_{j}} + \GVec{\tau}_l)
+ \psi_{B_{j-1}}(\GVec{\rho}_{A_{j}} + \GVec{\tau}_l)
\right] ,
\\
 \vare \psi_{B_{j}}(\GVec{\rho}_{B_{j}})
&=&
-\gamma_0 \sum_{l=1}^3  
\psi_{A_{j}}(\GVec{\rho}_{B_{j}} + \GVec{\tau}_l)
\nonumber\\
&&
\hspace{0mm}
+\gamma_1 \left[
\psi_{A_{j+1}}(\GVec{\rho}_{B_{j}})
+ \psi_{A_{j-1}}(\GVec{\rho}_{B_{j}})
\right] . \quad
\label{eq_sch1}
\end{eqnarray}
For even $j$,
\begin{eqnarray}
 \vare \psi_{A_{j}}(\GVec{\rho}_{A_{j}})
&=&
-\gamma_0 \sum_{l=1}^3  
\psi_{B_{j}}(\GVec{\rho}_{A_{j}} - \GVec{\tau}_l)
\nonumber\\
&&
\hspace{0mm}
+\gamma_1 \left[
\psi_{B_{j+1}}(\GVec{\rho}_{A_{j}})
+ \psi_{B_{j-1}}(\GVec{\rho}_{A_{j}})
\right] , \quad
\\
 \vare \psi_{B_{j}}(\GVec{\rho}_{B_{j}})
&=&
-\gamma_0 \sum_{l=1}^3  
\psi_{A_{j}}(\GVec{\rho}_{B_{j}} + \GVec{\tau}_l)
\nonumber\\
&&
\hspace{-20mm}
+\gamma_3 \sum_{l=1}^3  
\left[
\psi_{A_{j+1}}(\GVec{\rho}_{B_{j}} - \GVec{\tau}_l)
+ \psi_{A_{j-1}}(\GVec{\rho}_{B_{j}} - \GVec{\tau}_l)
\right] .
\label{eq_sch2}
\end{eqnarray}
Here we 
introduced the vectors from B site to the nearest neighboring A sites
as $\GVec{\tau}_1 = a(0,1/\sqrt{3})$,
$\GVec{\tau}_2 = a(-1/2,-1/2\sqrt{3})$,
and $\GVec{\tau}_3 = a(1/2,-1/2\sqrt{3})$,
and we set 
$\psi_{A_0} = \psi_{B_0} = \psi_{A_{N+1}} = \psi_{B_{N+1}} = 0$.

The states around $K$ point can be expressed 
in terms of the slowly-varying envelope functions
$ F_{A_j}, F_{B_j}$ as
\begin{eqnarray}
\psi_{A_{j}}(\GVec{\rho}_{A_j}) 
&=& C_{A_j} e^{i\Vec{K}\cdot\GVec{\rho}_{A_j}} F_{A_j}(\GVec{\rho}_{A_j}),
\\
\psi_{B_{j}}(\GVec{\rho}_{B_j}) 
&=& C_{B_j} e^{i\Vec{K}\cdot\GVec{\rho}_{B_j}} F_{B_j}(\GVec{\rho}_{B_j}),
\label{eq_env}
\end{eqnarray}
where $\Vec{K} = (2\pi/a)(1/3,1/\sqrt{3})$,
and $C_{A_j}$, $C_{B_j}$ are phase factors defined by
\begin{eqnarray}
&& C_{A_j} = -\omega^{-1}, \quad C_{B_j} = 1
\quad (j{\rm: odd}), \\
&& C_{A_j} = 1, \quad C_{B_j} = -\omega 
\quad (j{\rm: even}),
\label{eq_C}
\end{eqnarray}
with $\omega = \exp(2\pi i/3)$.
When $\GVec{\tau}_l$ is much smaller than
the length scale of the envelope functions, we have
\begin{equation}
\psi_{X}(\GVec{\rho} \pm \GVec{\tau}_l) 
\approx e^{i\Vec{K}\cdot(\GVec{\rho} \pm \GVec{\tau}_l)} 
\left( 1 \pm \GVec{\tau}_l\cdot
\frac{\partial}{\partial\GVec{\rho}}
\right)
F_{X}(\GVec{\rho}),
\label{eq_approx}
\end{equation}
with $X = A_j$ or $B_j$.

By substituting Eq.\ (\ref{eq_env}) with (\ref{eq_approx}),
into Schr\"{o}dinger's equations (\ref{eq_sch1}) and (\ref{eq_sch2}),
we have for odd $j$,
\begin{eqnarray}
 \vare F_{A_j}(\GVec{\rho}) &=& 
  \gamma k_-  F_{B_j}(\GVec{\rho})
+ \gamma' k_+  
\left[
F_{B_{j-1}}(\GVec{\rho}) + F_{B_{j+1}}(\GVec{\rho})
\right]
\nonumber\\
 \vare F_{B_j}(\GVec{\rho}) &=& 
  \gamma k_+  F_{A_j}(\GVec{\rho})
+ \gamma_1 
\left[
F_{A_{j-1}}(\GVec{\rho}) + F_{A_{j+1}}(\GVec{\rho})
\right], \nonumber\\
\end{eqnarray}
and for even $j$,
\begin{eqnarray}
 \vare F_{A_j}(\GVec{\rho}) &=& 
  \gamma k_-  F_{B_j}(\GVec{\rho})
+ \gamma_1 
\left[
F_{B_{j-1}}(\GVec{\rho}) + F_{B_{j+1}}(\GVec{\rho})
\right] ,
\nonumber\\
 \vare F_{B_j}(\GVec{\rho}) &=& 
  \gamma k_+  F_{A_j}(\GVec{\rho})
+ \gamma' k_-  
\left[
F_{A_{j-1}}(\GVec{\rho}) + F_{A_{j+1}}(\GVec{\rho})
\right],\nonumber\\
\end{eqnarray}
where $k_\pm = k_x \pm i k_y$ with
$k_x = \frac{1}{i}\frac{\partial}{\partial x}$,
$k_y = \frac{1}{i}\frac{\partial}{\partial y}$.
We also used an identity
\begin{eqnarray}
\sum_{l=1}^3 e^{-i\Vec{K}\cdot\GVec{\tau}_l}
\left(1\,\,\,\,\tau_l^x \,\,\,\, \tau_l^y
\right)
=
\frac{\sqrt{3}}{2}a\omega^{-1} (0\,\,\,\,i\,\,\,\, 1).
\end{eqnarray}
If we rewrite this set of equations into the matrix form
for a vector $(F_{A_1},F_{B_1},F_{A_2},F_{B_2},\cdots)$,
we finally obtain the Hamiltonian matrix (\ref{eq_H}).

The effective Hamiltonian for another valley $\Vec{K}' = (2\pi/a)(2/3,0)$
can be derived in a parallel way, while $\omega$ and $\omega^{-1}$
are exchanged in Eq.\ (\ref{eq_C}).

\section{Susceptibility formula}
\label{sec_app_b}

The susceptibility formula (\ref{eq_chi}) with (\ref{eq_F}) has been derived for the Luttinger-Kohn representation of the Bloch function\cite{Luttinger_and_Kohn_1955a} and therefore in systems described by the Hamiltonian consisting of the free electron kinetic energy $\hbar^2\hat{\bf k}^2/2m$ and terms linear in $\hat{\bf k}$.\cite{Fuku71}
We derive here the susceptibility formula which is valid in the general
Hamiltonian ${\cal H}(\hat{\Vec{k}})$ which includes $k$-square terms in off-diagonal matrix elements as well as $k$-linear terms.
We shall confine ourselves to the case without electron-electron interaction for simplicity.
\par
%
Consider the system described by the Schr\"odinger equation
\begin{equation}
{\cal H}\Big(\hat{\mb k}\!+\!{e \over \hbar} {\mb A}({\mb r}),\, {\mb r}\Big) \psi_\alpha({\mb r}) = \varepsilon_\alpha \psi_\alpha({\mb r}) ,
\end{equation}
with $\hat{\mb k}\!=\!-i\nabla$ and ${\mb A}$ being the vector potential.
The thermodynamic function $\Omega$ is given by
\begin{eqnarray}
\Omega &=& - k_{\rm B} T g_{\rm s} \frac{1}{V} \sum_\alpha 
\ln \big\{ 1 \!+\! \exp[\beta(\mu \!-\! \varepsilon_\alpha)] \big\} 
\nonumber\\ 
&=&
- k_{\rm B} T \, g_{\rm s} \frac{1}{V}  \int d\varepsilon \,
\Big(\!-\!{1\over\pi}\Big) {\rm Im} \, {\rm Tr} \, 
\frac{1}{\vare -{\cal H} + i0}
\nonumber\\
&&\qquad\qquad
\times\ln \big\{ 1 \!+\! \exp[\beta(\mu \!-\! \varepsilon)] \big\} , 
\label{eq_omega}
\end{eqnarray}
where $g_{\rm s}$ is the spin degeneracy and $V$ is the system volume.
\par
We consider an isotropic system and assume 
the vector potential ${\mb A}\!=\!(0,\, A)$, with
\begin{eqnarray}
 A(x) = { B \over 2 i q } ( e^{i q x} \!-\! e^{-i q x} ) ,
\quad
B(x) = B \cos(qx) ,
\end{eqnarray}
where we are going to take the long wavelength limit
$q\!\rightarrow\!0$, for which the field causes the response the same as
that due to a spatially uniform magnetic field.
In the presence of this
vector potential, the Hamiltonian changes from 
${\cal H}(\hat {\mb k})$
to ${\cal H}(\hat{\mb k}\!+\!\Delta{\mb k})$, with 
$\Delta{\mb k}\!=\!(0,\, \Delta k)$, where 
$\Delta k = (e/\hbar)A(x)$.
The Hamiltonian can be expanded as
\begin{equation}
 {\cal H}(\hat{\mb k}\!+\!\Delta{\mb k}) 
= {\cal H}_0 + {1\over 2} (
  \Delta k \, {\cal H}_y \!+\! {\cal H}_y \Delta k) + {1\over 2} (\Delta
  k)^2 {\cal H}_{yy},
\end{equation}
where
${\cal H}_0 \equiv {\cal H}({\mb k})$,
${\cal H}_y \equiv {\partial {\cal H}_0 / \partial \hat k_y }$, and
${\cal H}_{yy} \equiv {\partial^2 {\cal H}_0/ \partial \hat k_y^2 }$.
Note that in general $k$-square Hamiltonian
$\Delta k$ does not commute with ${\cal H}_y$ but
does with ${\cal H}_{yy}$.

Expanding the Hamiltonian 
up to the second order in the strength of the magnetic field $B$,
we have
\begin{eqnarray}
{\rm Tr} \Big( {1\over \varepsilon \!-\! {\cal H}} \!-\! {1\over \varepsilon \!-\! {\cal H}_0} \Big) & \!\!\! = \!\!\! & {1\over (2 i q l^2)^2} {\partial \over\partial\varepsilon} {\rm Tr} \Big( {1 \over \varepsilon \!-\! {\cal H}_0} {\cal H}_{yy} \nonumber \\
&& + {1 \over \varepsilon \!-\! {\cal H}_{{\mb q}/2}} {\cal H}_y {1 \over \varepsilon \!-\! {\cal H}_{-{\mb q}/2}} {\cal H}_y \Big) . \qquad
\label{eq_expand_in_B}
\end{eqnarray}
where $l=\sqrt{\hbar/eB}$ is the magnetic length,
${\cal H}_{\mb q} = {\cal H}(\hat{\mb k}+{\mb q})$,
$\Vec{q} = (q,0)$, and
we assumed that the system is translational invariant
(after the configuration average in the presence of impurities).

We then expand ${\cal H}_{\mb q}$ up to the second order in $q$,
to have
\begin{eqnarray}
&& \!\!\!\! {\rm Tr} \Big( {1\over \varepsilon \!-\! {\cal H}} \!-\!  
{1\over \varepsilon \!-\! {\cal H}_0 } \Big) \nonumber \\
&& \!\!\!\! = {1\over 16 l^4} 
{\partial \over\partial\varepsilon} 
{\rm Tr} \Big( 
G {\cal H}_x G {\cal H}_y G {\cal H}_x G {\cal H}_y 
-2 G {\cal H}_x G {\cal H}_x G {\cal H}_y G {\cal H}_y \nonumber \\
&& \qquad\qquad - \frac{1}{2} G {\cal H}_y G {\cal H}_{xx} G {\cal H}_y  
- \frac{1}{2} G {\cal H}_x G {\cal H}_{yy} G {\cal H}_x \Big),
\label{eq_trdiff}
\end{eqnarray}
with $G= (\vare-{\cal H}_0)^{-1}$.
This immediately gives the change of the thermodynamic potential
$\Delta\Omega = \Omega(B)-\Omega(0)$ with Eq.\ (\ref{eq_omega}).
The susceptibility $\chi$ is obtained by 
a relation
$\Delta\Omega = - (1/2) \chi \langle B(x)^2\rangle = 
(1/4)\chi B^2$ as
\begin{eqnarray}
 \chi &=& \frac{g_s}{4V}\frac{e^2}{\hbar^2}
\int d\vare f(\vare)\left(-\frac{1}{\pi}\right)\nonumber\\
&& 
\times {\rm Im}{\rm Tr}
 \Big( 
G {\cal H}_x G {\cal H}_y G {\cal H}_x G {\cal H}_y 
-2 G {\cal H}_x G {\cal H}_x G {\cal H}_y G {\cal H}_y
\cr 
&& 
- \frac{1}{2} G {\cal H}_y G {\cal H}_{xx} G {\cal H}_y  
- \frac{1}{2} G {\cal H}_x G {\cal H}_{yy} G {\cal H}_x \Big).
\end{eqnarray}
This gives Eq.\ (\ref{eq_chi}) with (\ref{eq_F_gen})
for the multilayer graphene.

When $\Delta k$ commutes with ${\cal H}_y$, we can simplify the formula
by noting in Eq.\ (\ref{eq_expand_in_B}) that
\begin{eqnarray}
&& {\rm Tr} \Big( 
{1 \over \varepsilon \!-\! {\cal H}_{{\mb q}/2}} 
{\cal H}_y 
{1 \over \varepsilon \!-\! {\cal H}_{-{\mb q}/2}} 
{\cal H}_y \Big) \nonumber \\
&& = {\rm Tr}\Big(  {1 \over \varepsilon \!-\! {\cal H}_{{\mb q}}} 
{\cal H}_y {1 \over \varepsilon \!-\! {\cal H}_0} {\cal H}_y \Big) 
\nonumber\\
&& = {\rm Tr}\Big(  {1 \over \varepsilon \!-\! {\cal H}_0} {\cal H}_y 
{1 \over \varepsilon \!-\! {\cal H}_{-{\mb q}}} {\cal H}_y \Big).
\end{eqnarray}
The susceptibility becomes
\begin{equation}
\chi = \frac{g_s}{2V}\frac{e^2}{\hbar^2}
\int d\vare f(\vare)\left(-\frac{1}{\pi}\right)
{\rm Im}{\rm Tr}
 \Big( 
G {\cal H}_x G {\cal H}_y G {\cal H}_x G {\cal H}_y 
\Big) , 
\end{equation}
giving Eq.\ (\ref{eq_chi}) with Eq.\ (\ref{eq_F}).



\begin{thebibliography}{99}


\bibitem{Novo04}
K. S. Novoselov, A. K. Geim, S. V. Morozov, D. Jiang, 
Y. Zhang, S. V. Dubonos, 
I. V. Grigorieva, and A. A. Firsov, Science {\bf 306}, 666 (2004). 


\bibitem{Novo05}
K. S. Novoselov, A. K. Geim, S. V. Morozov, D. Jiang, M. I. Katsnelson, 
I. V. Grigorieva, S. V. Dubonos, and A. A. Firsov, 
Nature {\bf 438}, 197 (2005).
 
\bibitem{Zhan05-2}
Y. Zhang, Y. W. Tan, H. L. Stormer, and P. Kim, Nature {\bf 438}, 201 (2005).


\bibitem{McCl56}
J. W. McClure, Phys. Rev. {\bf 104}, 666 (1956).

\bibitem{DiVincenzo_and_Mele_1984a}
D. P. DiVincenzo and E. J. Mele, Phys.\ Rev.\ B {\bf 29}, 1685 (1984).

\bibitem{Gonzalez_et_al_1994a}
J. Gonzalez, F. Guinea, and M. A. H. Vozmediano, Nucl.\ Phys.\ B {\bf 424}, 595 (1994).

\bibitem{Gonzalez_et_al_1996a}
J. Gonzalez, F. Guinea, and M. A. H. Vozmediano, Phys.\ Rev.\ Lett.\ {\bf 77}, 3589 (1996).

\bibitem{Shon_and_Ando_1998a}
N. H. Shon and T. Ando, J. Phys.\ Soc.\ Jpn.\ {\bf 67}, 2421 (1998).

\bibitem{Gonzalez_et_al_2001a}
J. Gonzalez, F. Guinea, and M. A. H. Vozmediano, Phys.\ Rev.\ B {\bf 63}, 134421 (2001).

\bibitem{Zheng_and_Ando_2002a}
Y. Zheng and T. Ando, Phys.\ Rev.\ B {\bf 65}, 245420 (2002).

\bibitem{Ando_et_al_2002a}
T. Ando, Y. Zheng, and H. Suzuura, J. Phys.\ Soc.\ Jpn.\ {\bf 71}, 1318 (2002).

\bibitem{Suzuura_and_Ando_2002b}
H. Suzuura and T. Ando, Phys.\ Rev.\ Lett.\ {\bf 89}, 266603 (2002).

\bibitem{Ando_2005a}
T. Ando, J. Phys.\ Soc.\ Jpn.\ {\bf 74}, 777 (2005).


\bibitem{Novo06}
K. S. Novoselov, E. McCann, S. V. Morozov, V. I. Falko,
M. I. Katsnelson, U. Zeitler, D. Jiang, F. Schedin, and
A. K. Geim, Nat. Phys. {\bf 2}, 177 (2006).

\bibitem{Ohta}
T. Ohta, A. Bostwick, T. Seyller, K. Horn, and E. Rotenberg, Science
{\bf 313}, 951 (2006);
T. Ohta, A. Bostwick, J. L. McChesney, T. Seyller, K. Horn, and E. Rotenberg,
Phys. Rev. Lett. {\bf 98}, 206802 (2007).


\bibitem{McCa}
E. McCann and V. I. Fal'ko, Phys. Rev. Lett. {\bf 96}, 086805 (2006).

\bibitem{McCa_gap}
E. McCann, Phys. Rev. B {\bf 74}, 161403(R) (2006).

\bibitem{Koshino_and_Ando_2006a}
M. Koshino and T. Ando, Phys.\ Rev.\ B {\bf 73}, 245403 (2006).

\bibitem{Nils_multi}
J. Nilsson, A. H. Castro Neto, F. Guinea, and N. M. R. Peres,
Phys. Rev. Lett. {\bf 97}, 266801 (2006).

\bibitem{Nils_imps}
J. Nilsson and A. H. Castro Neto,
Phys. Rev. Lett. {\bf 98}, 126801 (2007).


\bibitem{Guin}
F. Guinea, A. H. Castro Neto, and N. M. R. Peres,
Phys. Rev. B {\bf 73}, 245426 (2006).

\bibitem{Lati}
S. Latil and L. Henrard, Phys. Rev. Lett. {\bf 97}, 036803 (2006).

\bibitem{Part}
B. Partoens and F. M. Peeters,
Phys. Rev. B {\bf 74}, 075404 (2006).

\bibitem{Liu}
C. L. Lu, C. P. Chang, Y. C. Huang, J. M. Lu, C. C. Hwang, 
and M. F. Lin, J. Phys. Condens. Matter {\bf 18}, 5849 (2006).



\bibitem{McCl57}
J. W. McClure, Phys. Rev. {\bf 108}, 612 (1957);
{\it ibid}, {\bf 119}, 606 (1960).

\bibitem{Shar}
M. P. Sharma, L. G. Johnson, and J. W. McClure, Phys. Rev. B {\bf 9}, 
2467 (1974).

\bibitem{Safr}
S. A. Safran and F. J. DiSalvo, Phys. Rev. B {\bf 20}, 4889 (1979).

\bibitem{Safr84}
S. A. Safran, Phys. Rev. B {\bf 30}, 421 (1984).

\bibitem{Sait}
R. Saito and H. Kamimura, Phys. Rev. B {\bf 33}, 7218 (1986).

\bibitem{Gusy}
S. G. Sharapov, V. P. Gusynin, and H. Beck, Phys. Rev. B 
{\bf 69}, 075104 (2004).

\bibitem{Fuku07}
H. Fukuyama, J. Phys. Soc. Jpn. {\bf 76}, 043711 (2007).

\bibitem{Kosh}
M. Koshino and T. Ando, Phys. Rev. B {\bf 75}, 235333 (2007).


\bibitem{Wall}
P. R. Wallace, Phys. Rev. {\bf 71}, 622 (1947)

\bibitem{Slon}
J. C. Slonczewski and P. R. Weiss, Phys. Rev. {\bf 109}, 272 (1958).

\bibitem{Dres}
G. Dresselhaus and M. S. Dresselhaus,
Phys. Rev. {\bf 140}, A401 (1965).

%

\bibitem{Ajik}
H. Ajiki and T. Ando,
J. Phys. Soc. Jpn. {\bf 62}, 1255 (1993).


\bibitem{Kane_and_Mele_1997a}
C. L. Kane and E. J. Mele, Phys.\ Rev.\ Lett.\ {\bf 78}, 1932 (1997).

%

\bibitem{Char}
J. -C. Charlier, X. Gonze, and J. -P. Michenaud,
Phys. Rev. B {\bf 43}, 4579 (1991).



\bibitem{Toy}
W. W. Toy, M. S. Dresselhaus, and G. Dresselhaus,
Phys. Rev. B {\bf 15}, 4077 (1977).

\bibitem{Misu}
A. Misu, E. Mendez, and M. S. Dresselhaus, J. Phys. Soc. Jpn. 
{\bf 47}, 199 (1979).

\bibitem{Doez}
R. E. Doezema, W. R. Datars, H. Schaber, and A. Van Schyndel, 
Phys. Rev. B {\bf 19}, 4224 (1979).


\bibitem{Cast}
E. V. Castro, K. S. Novoselov, S. V. Morozov, N. M. R. Peres,
J. M. B. Lopes dos Santos, J. Nilsson, F. Guinea, A. K. Geim, and
A. H. Castro Neto,
cond-mat/0611342


\bibitem{Ohno}
T. Ohno and H. Kamimura, J. Phys. Soc. Jpn. {\bf 52}, 223 (1983).


\bibitem{Fuku71} 
H. Fukuyama, Prog. Theor. Phys. {\bf 45}, 704 (1971).


\bibitem{Fuku70}
H. Fukuyama and R. Kubo, J. Phys. Soc. Jpn. {\bf 28}, 570 (1970).

\bibitem{Naka}
M. Nakamura, cond-mat/0703355.


\bibitem{Stor}
H. L. Stormer, T. Haavasoja, V. Narayanamurti, A. C. Gossard, and W.
Wiegmann, J. Vac. Sci. Technol. B {\bf 1}, 423 (1983).

\bibitem{Mein}
I. Meinel, D. Grundler, S. Bargst\"adt-Franke, C. Heyn,
and D. Heitmann,  Appl. Phys. Lett. {\bf 70}, 3305 (1997)

\bibitem{Eise}
J. P. Eisenstein, H. L. Stormer, V. Narayanamurti, 
A. Y. Cho, A. C. Gossard, and C. W. Tu,
Phys. Rev. Lett. {\bf 55}, 875 (1985).

\bibitem{Pott}
A. Potts, R. Shepherd, W. G. Herrenden-Harker, M. Elliott, C. L. Jones,
A. Usher, G. A. C. Jones, D. A. Ritchie, E. H. Linfield, and M. Grimshaw,
J. Phys. C {\bf 8}, 5189 (1996).

\bibitem{Wieg}
S. A. J. Wiegers, M. Specht, L. P. Levy, M. Y. Simmons, 
D. A. Ritchie, A. Cavanna, B. Etienne, G. Martinez, and 
P. Wyder, Phys. Rev. Lett. {\bf 79}, 3238 (1997).

\bibitem{Luttinger_and_Kohn_1955a}
J. M. Luttinger and W. Kohn, Phys.\ Rev.\ {\bf 97}, 869 (1955).

\end{thebibliography}
\end{document}